\documentclass[twocolumn,showpacs,preprintnumbers,amsmath,amssymb,superscriptaddress,10pt,aps,prb]{revtex4-1}
\usepackage{epsfig,amsopn}
\usepackage{graphicx}
\usepackage{epstopdf}
\usepackage{sidecap}
\usepackage{amsmath,amssymb}
\usepackage{amsthm}
\usepackage{enumerate}

\newcommand{\angstrom}{\textup{\AA}}

\begin{document}
\title{$0$-$\pi$ transitions in a Josephson junction of irradiated Weyl semimetal}
\author{Udit Khanna}
\affiliation{Harish-Chandra Research Institute, Chhatnag Road, Jhunsi, Allahabad 211 019, India.}
\affiliation{Homi Bhabha National Institute, Training School Complex, Anushaktinagar, Mumbai, Maharashtra 400085, India}
\author{Sumathi Rao}
\affiliation{Harish-Chandra Research Institute, Chhatnag Road, Jhunsi, Allahabad 211 019, India.}
\affiliation{Homi Bhabha National Institute, Training School Complex, Anushaktinagar, Mumbai, Maharashtra 400085, India}
\author{Arijit Kundu}
\affiliation{Department of Physics, Indian Institute of Technology - Kanpur, Kanpur 208 016, India}
\begin{abstract}

We propose a setup for the  experimental realization of  anisotropic $0$-$\pi$ transitions of the Josephson current,
in a  junction whose link is made of irradiated Weyl semi-metal (WSM), due to the presence of chiral nodes.
The Josephson current through a time-reversal symmetric WSM has anisotropic (with respect to the orientation of the chiral nodes)
 periodic oscillations as a function of $k_0L$,  where $k_0$ is the (relevant) separation of the chiral nodes and $L$ is the length of the sample.
We then show that the  effective value of $k_0$ can be tuned with precision by irradiating the sample with linearly polarized light, which
does not break time-reversal invariance, 
resulting in $0$-$\pi$ transitions of the critical current. 
We also discuss the feasibility and robustness of our setup.

\end{abstract} 

\maketitle
\emph{Introduction.}---%
Weyl semimetals are 3D topological systems with two or more `Weyl' nodes in the bulk where valence and conduction bands touch~\cite{Vishwanath2011,Burkov2011a,Burkov2011b,Zyuzin2012a,Hosur2012}. 
According to a no-go theorem~\cite{Nielsen1981}, such Weyl nodes appear as pairs in momentum space with each of the nodes having a definite `chirality', a quantum number that depends on the Berry flux enclosed by a closed surface around the node. Attempts to understand the effects of such chiral nodes have  initiated an extensive field of research in the  last few years, both in theory as well as in experiments. Exotic transport phenomena have been predicted due to the presence of the chiral nodes~\cite{Vazifeh2013,Son2013,Turner2013,Biswas2013,Hosur2013,Burkov2014,Gorbar2014,Uchida2014, Khanna2014,Ominato2014,Sbierski2014,Burkov2015a,Burkov2015b,Goswami2015,Baum2015, Khanna2016,Behrends2016,Rao2016,Baireuther2016a,Tao2016,Marra2016,Li2016,Baireuther2016b,Madsen2016}  and an ever increasing number of experiments are being reported regularly~\cite{Xu2015a,Xu2015b,Lv2015a,Lv2015b,Lu2015,Jia2016} to confirm some of these predictions.

One such phenomenon, predicted recently, occurs at the interface of a WSM and a superconductor (SC), where both the processes, normal reflection and Andreev reflection, become inter-nodal~\cite{Uchida2014}, i.e, occur from one node to another node of opposite chirality. This extra transfer of momentum gives rise to an unusual oscillation~\cite{Khanna2016} of Josephson current in a SC-WSM-SC setup with the period being proportional to both the distance between the two nodes in momentum space and the length of the WSM sample. The observation of such oscillations could be a direct proof of the existence of the chiral nodes, but since  the momentum separation of the chiral nodes remains  fixed for a given material, it is not likely to be usable as a tuning parameter. On the other hand, it has also been understood recently how time periodic perturbations can affect the WSM~\cite{Wang2014,Hubener2016,Ishizuka2016,Chan2016,Yan2016,Deb2017}, and in particular, how a high
frequency incident elliptically polarized light can slightly modify the position of the effective chiral nodes of the WSM~\cite{Chen2016}. This provides the possibility that the external parameters controlling the perturbation can serve as tuning parameters in adjusting the separation of the chiral nodes.

\begin{figure}[!ht]
\centering
\includegraphics[width=0.45\textwidth]{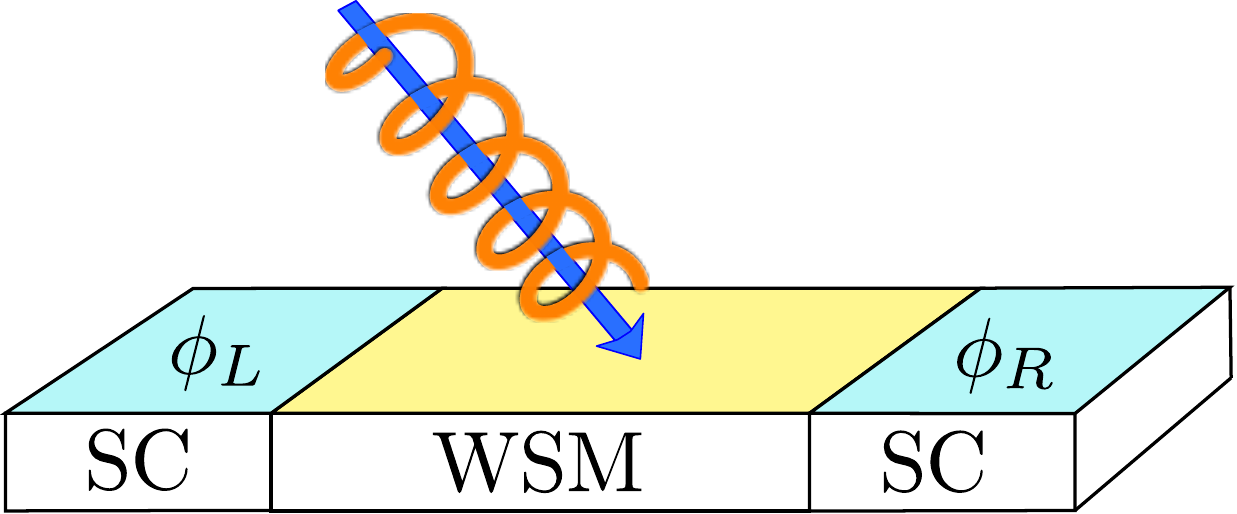}
\caption{ (Color online) The setup consists of a Josephson junction where a Weyl semimetal (WSM) has been sandwiched between two s-wave superconductors and is subjected to  a time-periodic perturbation with frequency higher than other relevant energy scales of the problem.}
\label{fig:fig1}
\end{figure}
\begin{figure*}[!ht]
\centering
\includegraphics[width=1.0\textwidth]{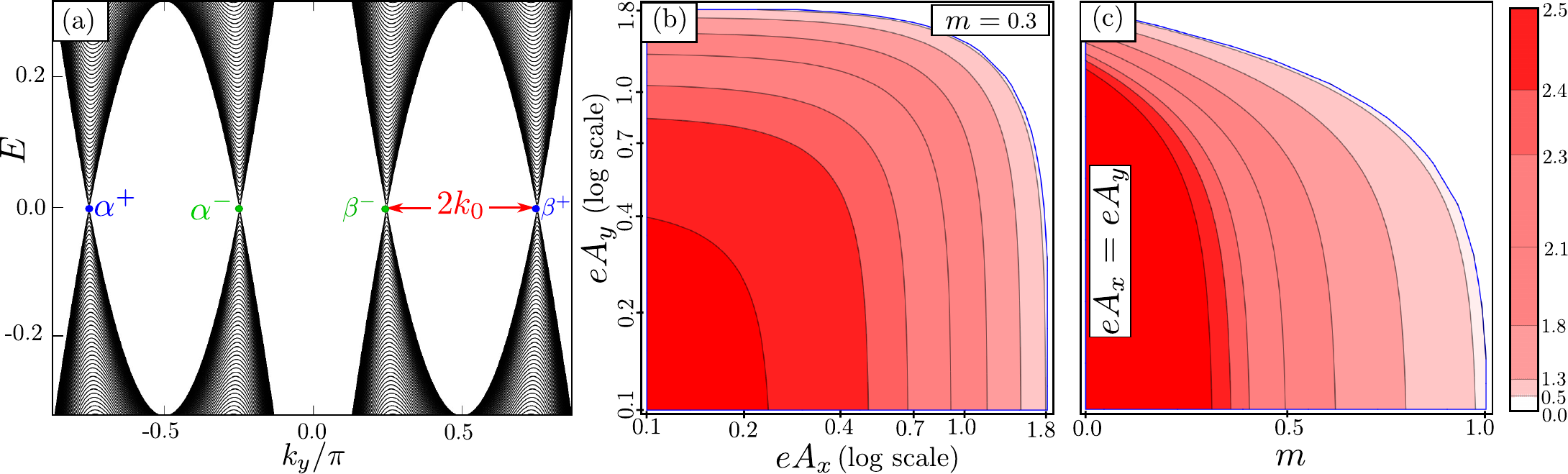}
\caption{ (Color online) (a) The low energy band structure of the model in Eq.~\ref{eq:H0}, which shows four Weyl nodes $\alpha^{\pm}$ and $\beta^{\pm}$ where +ve and -ve represents their respective chiralities. Each pair of nodes of opposite chiralities are separated by a distance in the momentum space by $2k_0=\pi - 2\sin^{-1}(m/\lambda)$. In the presence of  linearly  polarized light, the separation gets modified to $2\tilde{k}_0 = \pi - 2\sin^{-1}(m+1 - J_0(eA_x))/\lambda J_0(eA_y)$. The variation of $2\tilde{k}_0$ has been shown in (b) and (c) where $m/\lambda$=0.3 in (b), and $eA_x=eA_y$ in (c).}
\label{fig:fig2}
\end{figure*}

In this  paper,  we combine these two ideas in proposing a setup (see Fig.~\ref{fig:fig1}) for observing the unusual oscillation  in the Josephson effect in a WSM sandwiched between two superconductors by tuning only external parameters, such as the intensity or the phase of the polarized light, which impinges on the WSM.  As the period of oscillation of the Josephson current is proportional to both the length of the sample as well as the separation of the Weyl nodes, depending on the length of the sample, even a small  perturbative change in the separation of the nodes can give rise to a complete $0$-$\pi$ transition of the Josephson current. Note that the $0$-$\pi$ transition or change in sign of the critical current is expected~\cite{Bulaevski1977,Buzdin1982,Buzdin1991,Rmpsfs}  in time-reversal broken systems, like the SC-ferromagnet-SC junction and has even been observed experimentally~\cite{Ryazanov2001}. The transition  was also shown earlier in a time-reversal breaking WSM~\cite{Khanna2016}, but, motivated by the fact that almost all experimentally discovered WSMs~\cite{Xu2015a,Xu2015b,Lv2015a,Lv2015b,Lu2015,Jia2016} break inversion symmetry instead of time-reversal symmetry, in this paper, we show that it occurs in  a {\it time-reversal invariant} setup, using linearly polarized light. 

The existence of chiral nodes in a WSM is  also topologically protected  against perturbations, and,  although they can be moved around in momentum space, Gauss law prevents the annihilation of the nodes unless two of them with opposite chirality are brought together~\cite{Turner2013}. This provides  the robustness of our proposal.

\emph{Model and setup.}---%
Weyl semimetals  require either  time-reversal ($\mathcal{T}$) or inversion ($\mathcal{I}$) symmetry or both to be broken. 
The simplest model for a \textit{broken} $\mathcal{T}$ symmetric WSM has two Weyl nodes in the Brillouin zone, and has been studied extensively. But most of the present day WSM materials break  $\mathcal{I}$ symmetry. An inversion broken WSM is required to have at least four Weyl nodes in the Brillouin zone. A simple four band model with such an $\mathcal{I}$ broken WSM phase is the following~\cite{Chen2016}:
\begin{align}\label{eq:H0}
 \!\!\! H_0( {\bf k})= \sum_{i=x,y,z} \lambda_i \sigma^{i}\sin(k_i) + M({\bf k}) \tau^y\sigma^y + \epsilon({\bf k})\tau^y\sigma^x
\end{align}
where $\lambda_i$ are the spin orbit couplings, which are taken to be isotropic, 
i.e, $\lambda_i = \lambda$, $M(k) = m + 2 - \cos(k_x)-\cos(k_z)$ is the kinetic energy and $\epsilon(k) = \epsilon[1 - \cos(k_y)- \cos(k_z)]$ is a perturbation that breaks the $C_4$ symmetry about $k_y$ direction.
$\tau(\sigma)$ represents the orbital (spin) degree of freedom. $H_0({\bf k})$ is $\mathcal{T}$ symmetric, 
i.e, $\sigma^y H_0^{*}({\bf k})\sigma^y = H_0(-{\bf k})$, however, $\tau^x H_0({\bf k})\tau^x \neq H_0(-{\bf k} )$  for any value of the parameters due to the $M({\bf k})$ term and therefore inversion symmetry is intrinsically broken in this model. At $\epsilon=0$ this model realizes the WSM phase 
when $m<\lambda$, with four Weyl nodes along the $k_y$ axis at $k_y = \pm (\pi/2 \pm k_0)$ where $k_0 = \pi/2 - \sin^{-1}(m/\lambda)$ as shown in Fig.~\ref{fig:fig2}(a). The effective anisotropic 
Weyl Hamiltonian near these points is given by 
\begin{align}
 H_{\rm W} \approx \lambda\left[ \sigma^xk_x + \sigma^zk_z \pm \sigma^yk_y\frac{\sqrt{\lambda^2-m^2}}{\lambda} \right] + \mathcal{O}(k^2).\nonumber
\end{align}
Note that the anisotropy is controlled by the ratio of  $m/\lambda$. 
At $\epsilon \neq 0$ the $C_4$ symmetry about $k_y$ is absent and the Weyl nodes 
can move away from the $k_y$ axis in the $k_x-k_y$ plane. Further details of the model are discussed in the supplementary~\cite{supp} section.

In the presence of elliptically polarized light propagating in the $z$ direction, the Hamiltonian changes via the Peierls substitution 
${\bf{k} }\rightarrow {\bf k} + e{\bf A}(t)$, with,
\begin{align}
 {\bf A}(t) = (A_x\cos(\omega t), A_y\sin(\omega t + \theta), 0 ). 
\end{align}
At large (compared to the band width) driving frequency $\omega$, the system can still be effectively described by a static Hamiltonian. 
There are a number of approximation schemes~\cite{Mikami2016,Feldman1984,Mananga2011,Casas2001,Kuwahara2016,Eckardt2015,Bukov2015} available to find the effective Hamiltonian. In  the van Vleck approximation~\cite{Eckardt2015,Bukov2015}, the effective Hamiltonian to order $1/\omega$,  is given by
\begin{align}
 H_{\rm eff} = H_{(0)} + \sum_{n\neq 0} \frac{H_{(n)} H_{(-n)}}{n\omega} + \mathcal{O}(1/\omega^2),
\end{align}
where $H_{(n)}$ are the Fourier components of the time dependent Hamiltonian. 
The Fourier components can be found analytically~\cite{supp} and 
the effective Hamiltonian $H_{\rm eff}({\bf k}) = \tilde{H}_0({\bf k}) + H'({\bf k})$ 
where $\tilde{H}_0$ is equal to the bare Hamiltonian in absence of light $H_0({\bf k})$ 
with anisotropic renormalization of the parameters : $\lambda_i = \lambda J_0(eA_i)$ and 
\begin{align}
 M(k) &= m + 2 - \cos(k_x)J_0(eA_x)-\cos(k_z),\nonumber \\ {\rm and} ~~
 \epsilon(k) &= \epsilon [1 - \cos(k_y)J_0(eA_y)- \cos(k_z)].
\end{align}
The additional term is $H' = D_3\sigma_Z$ with, 
\begin{align}
 D_3 =& \sum_{n=1}^{\infty}\frac{4}{n\omega} J_n(eA_x)J_n(eA_y)\sin\left(n\theta + \frac{n\pi}{2}\right)\times \nonumber\\
 & \left\{ \begin{array}{ll}
           \lambda^2\sin k_x\sin k_y - \epsilon \cos k_x \cos k_y, & \text{if}~n~\text{is even}\\
           -\lambda^2\cos k_x\cos k_y + \epsilon \sin k_x \sin k_y,  & \text{if}~n~\text{is odd}.\\
          \end{array} \right. \nonumber
\end{align}

\noindent
For small amplitudes (when $eA_i \ll 1$), this additional term can be neglected. Further, for linearly polarized light, $\theta = \pi/2$, the additional term vanishes. This is the case that we will consider in the rest of the paper.

The position of Weyl nodes in the irradiated WSM, described by $H_{\rm eff}$, can now be controlled by the amplitude of the incident radiation. At $\epsilon = 0$, the four Weyl nodes are along the $k_y$ axis at $k_y = \pm (\pi/2 \pm \tilde{k}_0)$ where $\tilde{k}_0 = \pi/2 - \sin^{-1}(|m_{\rm eff}|/\lambda_{\rm eff})$. The material dependent parameters $m$ and $\lambda$, which cannot be 
directly tuned easily, change to effective values given by $m_{\rm eff} = (m+1-J_0(e A_x))$ and 
$\lambda_{\rm eff} = \lambda J_0(e A_y)$. The separation of two nearby Weyl nodes with opposite chirality is now given by $2\tilde{k}_0$. For small amplitudes $eA_i$, the change in separation is,
\begin{align}\label{eq:delk}
 2(\tilde{k}_0 - k_0) = 2\delta k_0 \approx - \frac{(eA_x)^2 + m(eA_y)^2}{2 \sqrt{\lambda^2-m^2}}.
\end{align}
We plot $2\tilde{k}_0$ in  Figs.~\ref{fig:fig2}(b) and ~\ref{fig:fig2}(c) with the variation 
of $eA_x$ and $eA_y$ of the incident radiation amplitude and with $m$ 
and $eA_0 = |eA_x| = |eA_y|$ respectively, demonstrating the 
tunability of the separation of the Weyl nodes by incident linearly polarized light.

\emph{Josephson current.}---%
Andreev reflection in a WSM-SC system can take place involving various nodes, $\alpha^{\pm}, \beta^{\pm}$, as shown in Fig.~\ref{fig:fig2}(a). If there was no chirality associated with the nodes, one would naively expect pairing between nodes $\alpha^{+(-)}$ to $\beta^{+(-)}$, which are the zero-momentum pairings. But, due to the overall spin conserving processes at a WSM-superconductor (SC) junction~\cite{Uchida2014,Khanna2016}, the helical quasiparticle excitations at the Weyl nodes allow only transport between nodes of opposite chirality. This implies that inter-nodal scattering can occur, say, from node $\alpha^{+}$ to node $\alpha^{-}$ (an inter-nodal distance of $2k_0$) or from node $\alpha^{+}$ to $\beta^{-}$, (an inter-nodal distance of $\pi$). As shown, for $\mathcal{T}$ broken systems, in Ref.~\cite{Khanna2016}, such momentum separation of the chiral nodes, $2k_0$ contributes to transfer of momentum at the superconducting interface each time a reflection/Andreev reflection process takes place. As a result, the energies of the bound-states between the two superconductors oscillate as a function of $k_0L$ with oscillation frequency of $\pi$.

\begin{figure}
\centering
\includegraphics[width=0.48\textwidth]{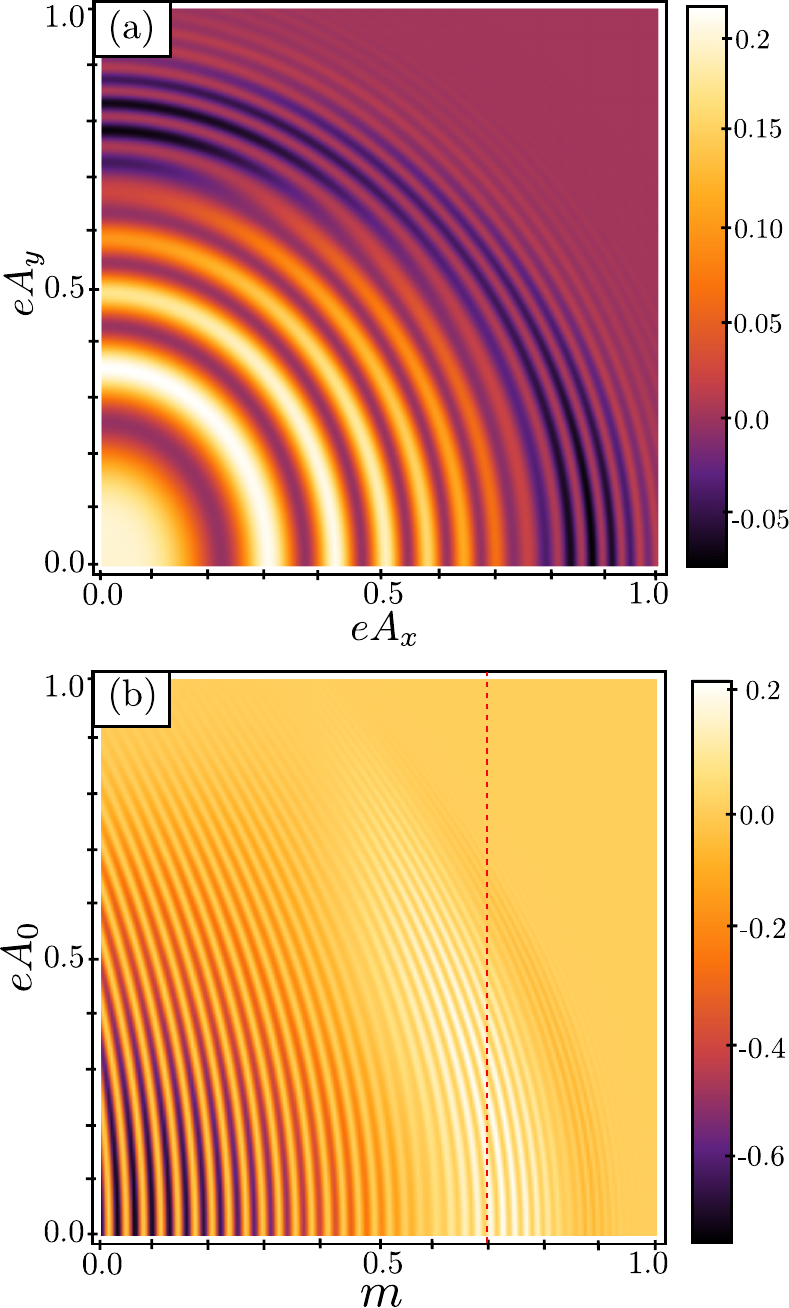}
\caption{ (Color online) Variation of the Josephson current at the phase difference $\phi = \pi/2$ in the parameter space of our model. (a) The external parameters $eA_x$ and $eA_y$, which control the plane of polarization, provide excellent tunability to the $0$-$\pi$ transition. In (b) we show how varying the amplitude of the incident field $eA_x=eA_y=eA_0$ gives rise to the $0$-$\pi$ transitions. The vertical line in (b) has been further emphasized in Fig.~\ref{fig:tune}. 
The current $J$ is in units of $e \lambda/ \hbar a$ ($a = 1$ is the lattice constant) and the legend shows the value of $J\times 100$. The 
parameters used are $\lambda = 1$, $\omega = 20$, $L = 70$ and in (a) $m = 0.7$. }
\label{fig:JvA}
\end{figure}
Although our model does not break time reversal, the main conclusions of Ref.~\cite{Khanna2016} do carry over to our system as well because these oscillations are the manifestation of chiral nodes rather than symmetry. We can now expect two possible momentum scales ($\pi$ and 2$k_0$); however, as we shall see below, the smaller of the two momentum scales, $2k_0$ is the relevant momentum scale 
at low energy transport. The expected oscillation in Josephson current results from the fact that the Josephson current is carried by these bound states.

We proceed to numerically evaluate Josephson current using a Green's function technique~\cite{Vidal1994}, further details of which are given in the supplemental material~\cite{supp}. The numerical method is well tested, and for various parameter ranges, we make sure that the continuum contribution to the Josephson current remains small compared to the bound-state contribution. In our numerical simulation we take a superconducting pair potential $\Delta = 0.05$ times the hopping amplitude in the Weyl semimetal. A large system size is required to ensure that the finite size gap in the WSM is smaller than the superconducting gap. One way we check this is to reach at least the length $L$ after which the current decays as $1/L$.  We also restrict ourselves to $\epsilon=0$, so that  the Weyl nodes are along the $y$-axis.

First, we study the Josephson current in the absence of light. As we increase the length $L$, we find oscillations in $J_y$  (the Josephson current in 
the direction of the Weyl node splitting  in momentum space) with a period of oscillation $\pi/k_0$. In contrast, the Josephson current along the perpendicular $x$-direction, $J_x$ is independent of $L$, apart from the trivial $1/L$ fall off. As has already been stated, the effect of irradiating the WSM sample by linearly polarized light is to change the effective distance between the Weyl nodes. This leads to (anisotropic) oscillations in the Josephson current as a function of the amplitudes of the impinging light. The variation in $J_y$ as a function of the amplitudes of linearly polarized light is shown in Fig.~\ref{fig:JvA}(a), where the frequency of the drive, $\omega$, is much larger than the band-width. Note that the oscillations are not quite radially symmetric, which is not unexpected, since the change in $k_0$ is not symmetrically affected, c.f, Eq.~(\ref{eq:delk}). The alternation of the positive and negative values of the current or the $0$-$\pi$ oscillations are clearly visible and can be further tuned by changing the amplitudes $eA_x$ and $eA_y$. Fig.~\ref{fig:JvA}(b) shows  the oscillations in the parameter space of $m$ and the amplitude of the incident light $eA_0 = eA_x =e A_y$. It is interesting to note that the oscillations in the current (Fig.~\ref{fig:JvA}) roughly match the graph of the change in the  momentum distance between the nodes  (Fig.~\ref{fig:fig2}) for the same changes in parameters. This confirms our claim that the oscillations that are seen in the Josephson current are essentially oscillations in $\tilde{k}_0L$.

In passing we would like to point out that, the Josephson current, in general, can oscillate with other system parameters. Such oscillations may appear, among other reasons, due to modifications of density of states, although $0$-$\pi$ transitions are unlikely. Moreover, such oscillations would not depend on the size of the system in the limit of large system size. We briefly discuss such variations of the Josephson current, $J_x$, with radiation parameters in the supplemental material~\cite{supp}.

\begin{figure}[t]
\centering
\includegraphics[width=0.48\textwidth]{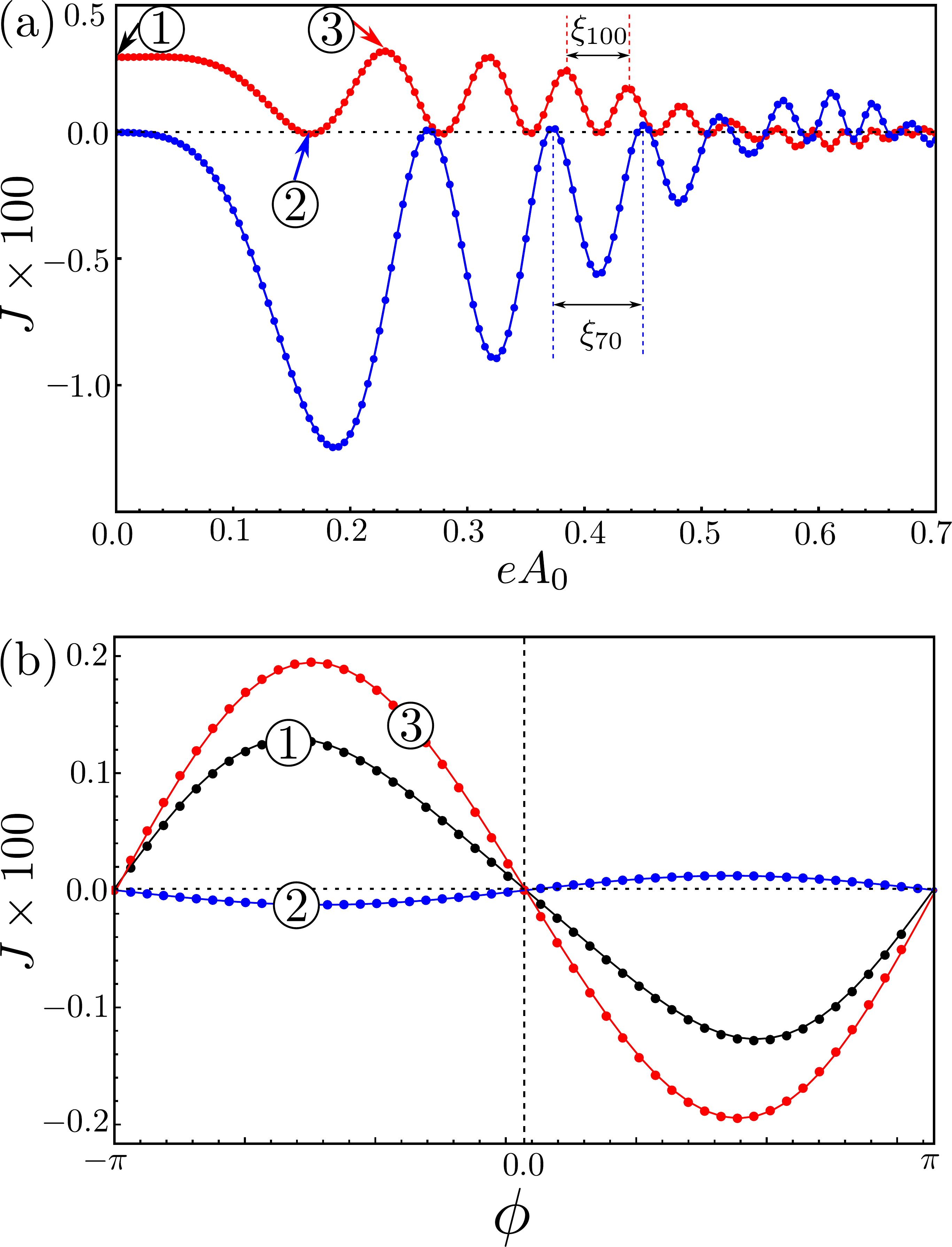}
\caption{ (Color online) Oscillations of the Josephson current   $J$ ($\times 100$),  in units of $e \lambda / \hbar a$, and the $0$-$\pi$ transition in the WSM as a function of (a) the amplitude of the drive $eA_x=eA_y=eA_0$ (the red line for length $L=100$ and the blue line for $L=70$ in units of the lattice constant $a$) and (b) the phase difference of the superconductors 
(the lines labelled $1$, $2$ and $3$ correspond to $e A_0 = 0$, $0.16$ and $0.22$ respectively). The 
parameters used are $m = 0.7$, $\lambda = 1$ and $\omega = 20$. }
\label{fig:tune}
\end{figure}

\emph{Tunability of the $0$-$\pi$ transition.}---%
In Fig.~\ref{fig:tune}, we show the Josephson current for different values of the amplitude of incident light.  The point to note here is that even a small change in the amplitude of light can cause $0$-$\pi$ transitions in the critical current. 
This is the central result of the paper,  {\it $0$-$\pi$  transitions in the Josephson current can be tuned by irradiating a WSM sample}. For an already irradiated sample, only a small change in intensity is required to observe a  $0$-$\pi$ transition. The change in amplitude of $A_y$ required to observe one full oscillation, in the limit of large $L$ (the length of the WSM) and for linearly polarized light with  $A_x=0$ is,
\begin{align}\label{eq:delI}
 \xi_L \equiv e \delta A_y \approx \frac{\pi}{J_{1}(eA_y)} \frac{\sqrt{\lambda^2-m^2}}{m L}.
\end{align}
The larger the system size, the smaller is this change in amplitude required, $\xi_L\propto 1/L$. 
The intensity of the light is $I = c_0 A_y^2$, where $c_0 = \frac{1}{2} c \epsilon_0 \omega^2$, $c$ being the speed of light and $\epsilon_0$ being the dielectric constant.
The corresponding change in intensity required to observe a full oscillation is 
$\delta I = 2c_0A_y \delta A_y \approx 4\pi c_0 \sqrt{\lambda^2-m^2}/me^2 L $ 
for a small drive  amplitude.
In WSM candidate materials like TaAs, the average $v_f$ has been 
measured~\cite{Xu2015a,Xu2015b,Lv2015a,Lv2015b,Lu2015} to be 
$\hbar v_f = 2 eV \angstrom$ (at $300 K$) and the 
average band-gap ($2m$) at the $\Gamma$ point is $ \approx 0.2 eV $. So we approximate $\lambda =  \hbar v_F=2 eV \angstrom$ and 
$m = 0.1 eV$. Using the average lattice constant $ a = 5 \angstrom$, $\hbar \omega = 117 meV$ for 
a CO$_2$ laser with $e A_y = 0.1 \angstrom^{-1}$ and assuming the 
length of WSM to be $ 100 \mu m $, we find $\delta I \approx 2 \times 10^{10} W/m^2 = 10^{-3} I $.   Further, with appropriate choice of parameters,
the system can also act as an \textit{on-off switch}, where turning on the laser changes the sign of the current~\cite{supp}, which may be of technological significance. 

\emph{Discussion.}---%
A discussion of the shortcomings of our analysis and the conditions needed for the successful observation of the physics that have we described here is in order. A few approximations have been made in our analysis which may not  hold in a realistic sample. We have assumed that the system is uniformly irradiated by a coherent source. However, 
in a real experiment, the irradiation within the sample will be limited to be within the skin depth. For a skin depth of a few layers of the atomic structure, the actual value of the radiation needed to observe an oscillation, Eq.~\ref{eq:delI}, will need to be modified, although we expect the effect to remain intact. One advantage of our proposal is  that the radiation intensity required is low and in fact, decreases with increasing system size,  though the length of the system might be limited by the coherence length of the laser. We leave more detailed studies studies, including the effect of  decaying radiation amplitude through  the sample, for the future.

We have presented our results for  a simple model of a WSM  with four Weyl nodes along a particular axis. Real systems often have many more Weyl nodes. However, along any particular direction, it is not natural to expect more than four Weyl nodes, so we expect our results to hold even in those systems as long as the Josephson current is measured along the direction in which the Weyl nodes are expected.

\emph{Summary and Conclusion.}---%
To summarize, in this paper we have studied, first, how the  $0$-$\pi$ transitions in the Josephson current in a time-reversal invariant WSM can result from the presence of chiral nodes. Without breaking the time-reversal symmetry, 
and hence, retaining the topological stability of the Weyl nodes, we have presented a way to observe such oscillations by an all-electric tunable setup using  linearly polarized light. We have presented numerical evidence of such $0$-$\pi$ transitions,  which are  highly anisotropic and  depend strongly on the orientation of the Weyl nodes. 

\emph{Note.}---%
During the review of our manuscript, we noticed the work of Bovenzi {\textit{et al}~\cite{Beenakker1704}. 
They study the normal and Andreev reflection processes at the junction of a WSM 
(with broken $\mathcal{T}$ symmetry) with a normal superconductor and observe that,  
while reflection within the same node is always blocked,  
Andreev reflection from one node to another can also be blocked at a WS junction if 
the interface or pair potential does not couple the two chiralities. 
This extra blocking is labelled 
``chirality blockade'' in their work. 


\section*{Appendix}

\subsection{Lattice model of a WSM without inversion symmetry}

We consider a four band fermionic model on a cubic lattice that has multiple WSM phases with 
different numbers of Weyl nodes. 
Assuming periodic boundary conditions in all directions, the hamiltonian is 
$\hat{\mathcal{H}} = \sum_{\bf k} \psi^{\dagger}_{\bf k} H({\bf k}) \psi_{\bf k}$ where 
$\psi_{\bf k}$ is the four component electron annihilation operator and 
\begin{align}\label{eq:ham}
H({\bf k}) = \sum_{i=x,y,z} \frac{\lambda_i}{a}  \sigma^i \sin (k_i a) + M({\bf k}) \tau^y \sigma^y ~.
\end{align}
Here $M({\bf k}) = m + 2t_h \left[ 2 - \cos(k_x a) - \cos(k_z a) \right]$, 
$m$ is half the band gap at the $\Gamma$ point, $t_h$ is the nearest neighbour coupling in the $x$ and $z$ directions, 
$\lambda_i$ are the anisotropic spin orbit couplings and $a$ is the lattice constant. 
$\sigma$ ($\tau$) denote the spin (orbital) degree of freedom. 
This model has a $C_4$ rotational symmetry about the $k_y$ axis which can be lifted by adding a term 
$\epsilon({\bf k})\tau^y \sigma^x$ where 
$\epsilon({\bf k}) = \epsilon \left[ 1 - \cos(k_y a) - \cos(k_z a) \right]$. 
At $t_h = 0.5$ and $a = 1$, this yields Eq. (1) of the main text. 
For brevity, in this work we assume $m > 0$ and isotropic spin orbit terms : $\lambda_i = \lambda$. 
Further, we only consider the case $\epsilon = 0$.

The model satisfies $\sigma^y H^*({\bf k}) \sigma^y = H({\bf -k})$ and therefore it is 
time reversal invariant. However $\tau^x H({\bf k}) \tau^x \neq H({\bf -k})$, i.e. 
the model breaks inversion symmetry. Thus a WSM phase can be expected in the model.
The eigenvalues of $H({\bf k})$ are $\pm \sqrt{E_{{\bf k}\pm}}$ where,
\begin{align}\nonumber
E_{{\bf k}\pm} = \left( \frac{\lambda}{a} \right)^2 &\left[ \sin ^2(k_x a) +  \sin ^2(k_z a) \right] +\\ \nonumber 
&\left[ \left(\frac{\lambda}{a}\right) \sin (k_y a) \pm M({\bf k}) \right]^2 .
\end{align}
Then $E_{{\bf k}\pm} = 0$ at $k_x, k_z = 0, \pi/a$ and $\lambda \sin(k_y a) = \pm a M({\bf k})$. 
Expanding $H({\bf k})$ around any of these zeros gives 
an effective Weyl hamiltonian. Thus the model describes a WSM if the zeros of $E_{{\bf k}\pm}$ exist. 

\begin{figure}[ht]
  \includegraphics[width=0.4\textwidth]{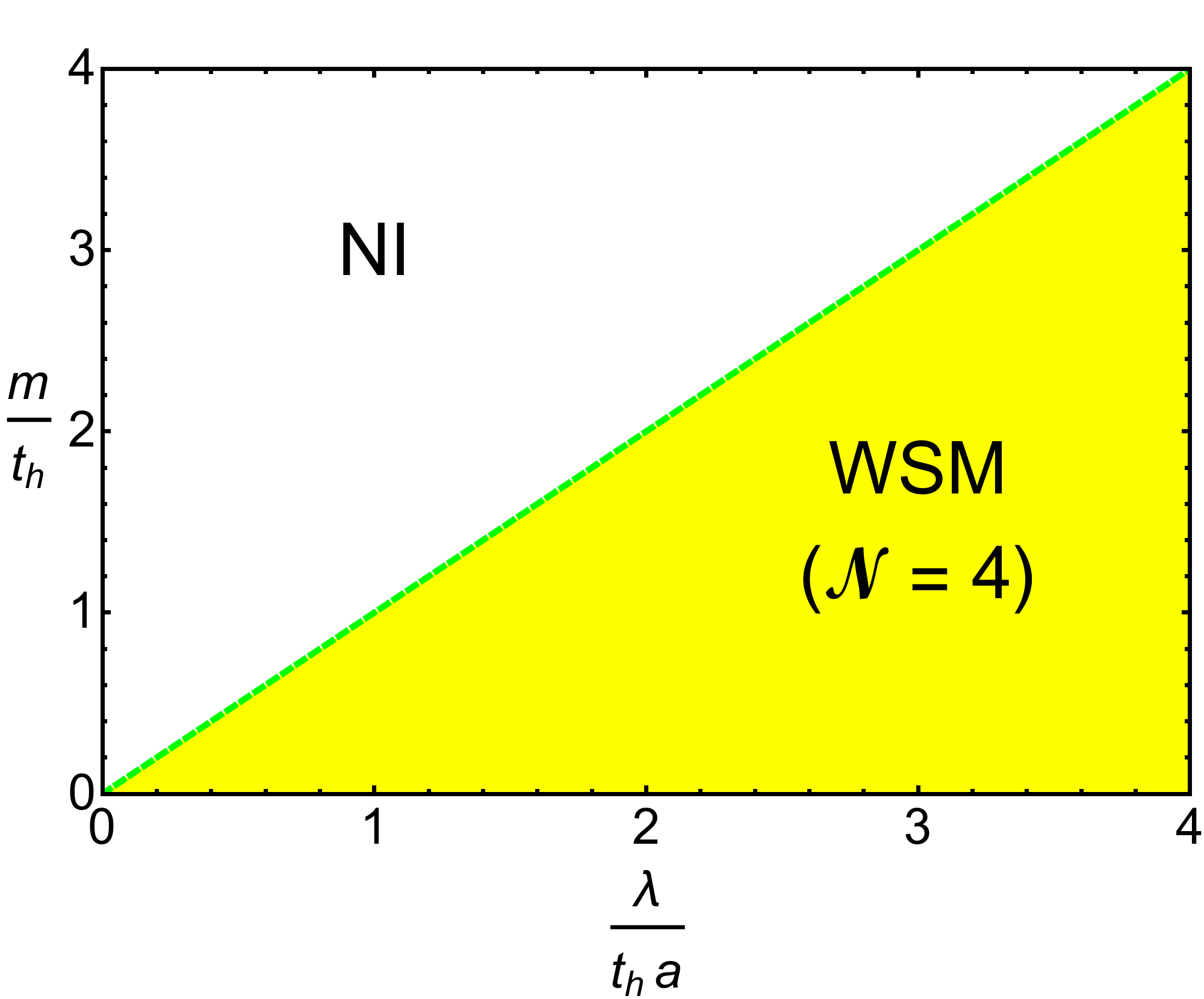}
	\caption{Phase diagram of the model showing the normal insulator and the minimal WSM phase at $\epsilon=0$.} \label{phasediagram}
\end{figure}

If $\lambda < ma$, then there are no zero energy states in the 
bandstructure and the model describes a normal insulator. 
At $\lambda = ma$, the bulk gap closes at ${\bf k_0}a = (0,\pm \pi/2,0)$ and 
close to these points $H({\bf k})$ is, (upto $O(q_x^2 + q_z^2)$)
\begin{align}\nonumber
H({\bf k_0 + q}) = \lambda \left[ \sigma^x q_x + \sigma^z q_z 
\mp \sigma^y \left( \frac{q_y^2 a}{2} \right) \right]
\end{align}
which is neither a Dirac nor a Weyl hamiltonian since it is linear in $q_x$ and $q_z$ but quadratic in $q_y$.

If $ma < \lambda < (m + 4t_h)a$, then the model has nodes at $k_x = 0 = k_z$, $\sin(k_y a) = \pm a m / \lambda$.  
The latter equation has 4 solutions : $\pm \sin^{-1}(am/ \lambda)$, $\pm(\pi - \sin^{-1}(am/ \lambda)$ and 
close to these nodes, the hamiltonian is, (upto $O({\bf q}^2)$)
\begin{align}\nonumber
H({\bf k_0 + q}) \approx \lambda \left[ \sigma^x q_x + \sigma^z q_z 
\pm \sigma^y \sqrt{1 - \frac{a^2m^2}{\lambda^2}} q_y \right]
\end{align}
which is the anisotropic Weyl hamiltonian with chirality $\pm 1$. 
Due to Kramer's theorem, the minimal model for inversion 
symmetry broken WSM must have atleast four Weyl nodes. 
Therefore this phase of the model describes the simplest possible WSM with broken inversion symmetry. 

Additional Weyl nodes can appear on the $k_x, k_z = \pi/a$ 
planes, at larger values of $\lambda$ : a total of 12 
for $(m + 4t_h)a < \lambda < (m + 8t_h)a$ and 16 for 
$ (m + 8t_h)a < \lambda $. 
In this work we use parameters so that only 4 nodes exist. Additional nodes are expected to increase the total current 
but not affect the results qualitatively.

\subsection{Effective Hamiltonian in the presence of polarised light}

In the presence of elliptically polarized light of frequency $\omega$ propagating along the $z$ direction, 
the system is described by a time dependent hamiltonian which is related to 
$H({\bf k})$ by the Peierls substitution. In SI units this means, 
$ {\bf k} \rightarrow {\bf k} + (e/{\hbar}) {\bf A}(t) $, where 
${\bf A}(t) = (A_x \cos (\omega t), A_y \sin (\omega t + \theta), 0) $.
The resulting time dependent Hamiltonian, can then  be written as a Fourier series
$ H({\bf k}, t) = \sum_{n} H_{(n)}({\bf k}) e^{i n \omega t} $. 
The Fourier modes $H_{(n)}({\bf k})$ are $4 \times 4$ matrices and can therefore 
be written as $H_{(n)}({\bf k}) = \sum_i d_{n,i} ({\bf k}) \zeta^{i}$ where we have defined matrices
$ \zeta^{i}$ in the spin ($\sigma$) and orbital ($\tau$) space as 
\begin{align}
  \zeta^1 &= I_\tau \sigma^x   \textbf{ , } \zeta^2 = I_\tau \sigma^y  \textbf{ , } \zeta^3 = I_\tau \sigma^z  \textbf{ , } \nonumber \\
  \zeta^4 &= \tau^y \sigma^x \textbf{ , } \zeta^5 = \tau^y \sigma^y \textbf{ , } \zeta^6 = \tau^y \sigma^z ~.
\end{align}
For brevity, we define $e^* = ea/\hbar$. Then,   
using some identities on Bessel functions we can compute 
the Fourier modes explicitly as, 
\begin{align}
  d_{n,1} &= \frac{\lambda}{a} J_{n}(e^* A_x) \sin (k_x a + \frac{n \pi}{2}),  \nonumber \\
  d_{n,2} &= \frac{\lambda}{a} J_{n}(e^* A_y)  e^{i n \theta} 
  \begin{cases}
    \sin (k_y a) & \text{if } n \text{ is even}\\
    -i \cos (k_y a) & \text{if } n \text{ is odd}\\
  \end{cases} \nonumber  \\
  d_{n,3} &= \frac{\lambda}{a} \sin (k_z a) \delta_{n, 0}, \nonumber  \\
  d_{n,4} &= \epsilon [1 - \cos(k_z a)] \delta_{n,0}   
  \nonumber \\ &~~~~~- \epsilon J_{n}(e^* A_y)  e^{i n \theta}  
  \begin{cases}
  \cos (k_y a) &  \text{if } n \text{ is even}\\
  i \sin (k_y a)  &\text{if } n \text{ is odd}\\
  \end{cases} \nonumber  \\
  d_{n,5} &= \left[ m + 2t_h(2 - \cos (k_z a)) \right] \delta_{n,0} 
  \nonumber \\ &~~~~~- 2t_h J_{n}(e^* A_x) \cos (k_x a + \frac{n \pi}{2}) ~~\text{ and }~~ \nonumber  \\
  d_{n,6} &= 0.
\end{align}

Assuming that the frequency of radiation $\omega$ is much larger than the band-width of the 
model, we can replace the time dependent hamiltonian $H({\bf k}, t)$ by an 
effective static hamiltonian called the Floquet hamiltonian $H_{\text{eff}} ({\bf k})$. 
In the van Vleck approximation\cite{Bukov2015} this is,
\begin{align} \nonumber
  H_{\text{eff}}({\bf k}) =  H_{(0)}({\bf k}) + \frac{1}{\hbar \omega} 
  \sum_{n \neq 0} \frac{H_{(n)}({\bf k}) H_{(-n)}({\bf k})}{n} + O(\frac{1}{\omega^2})
\end{align}
Using the expressions for $H_{(n)}({\bf k})$ defined above, we find 
$H_{\text{eff}}({\bf k}) = \sum_{j,a} \left[ D_{a}^{(j)}({\bf k}) 
/{(\hbar \omega)^j} \right] \zeta^{a}$. 
The tree level ($j = 0$) terms are - 
  \begin{align} 
  D_{1}^{(0)}({\bf k}) &= \frac{\lambda}{a} \sin(k_x a) J_0(e^* A_x), \nonumber \\ 
  D_{2}^{(0)}({\bf k}) &= \frac{\lambda}{a} \sin(k_y a) J_0(e^* A_y), \nonumber \\
  D_{3}^{(0)}({\bf k}) &= \frac{\lambda}{a} \sin(k_z a), \nonumber \\
  D_{4}^{(0)}({\bf k}) &= \epsilon \left[1 - \cos (k_z a) - \cos (k_y a) J_0(e^* A_y) \right], \nonumber \\
  D_{5}^{(0)}({\bf k}) &= m + 2t_h \left[2 - \cos (k_z a) - \cos (k_x a) J_0(e^* A_x) \right], \nonumber \\
  D_{6}^{(0)}({\bf k}) &= 0. 
  \end{align}
Clearly at the zeroth order, $H_{\text{eff}}({\bf k})$ is equal to 
the bare hamiltonian $H({\bf k})$ (Eq.~\ref{eq:ham}) with anisotropic renormalisation of the parameters. Therefore 
in the presence of light, the positions of the Weyl nodes change slightly.
The leading order ($j = 1$) terms are - 
\begin{align}  
  D_{1}^{(1)}({\bf k}) &= D_{2}^{(1)}({\bf k}) = 0, \nonumber \\
  D_{3}^{(1)}({\bf k}) &= \frac{4}{\hbar\omega} \sum_{n=1}^{\infty} 
  \frac{1}{n} J_n(e^* A_x) J_n(e^* A_y) \sin (n\theta + \frac{n\pi}{2}) \times\nonumber \\ 
  &\!\!\!\!\!\!\!\!\!\!\!\!\!\!\!\!\!\!\!  \begin{cases}
  (\frac{\lambda}{a})^2 \sin (k_x a) \sin (k_y a) - 2 t_h\epsilon \cos (k_x a) \cos(k_y a)  & \text{if } n \text{ is even}\\
    -(\frac{\lambda}{a})^2 \cos (k_x a) \cos (k_y a) + 2 t_h\epsilon \sin (k_x a) \sin (k_y a) & \text{if } n \text{ is odd}\\ 
  \end{cases}  \nonumber  \\
  D_{4}^{(1)}({\bf k}) &= D_{5}^{(1)}({\bf k}) = D_{6}^{(1)}({\bf k}) = 0. 
\end{align}
The first order correction to the effective hamiltonian is of the form 
$D_{3} ({\bf k}) \sigma^z$ and has the effect of moving the Weyl nodes 
in the $k_z$ direction so that they are not all in the same plane. 
Here we consider linearly polarized light ($\theta = \pi/2$), so that this 
correction vanishes exactly and the Weyl nodes remain fixed on the $k_z = 0$ plane. 

Since $H_{\text{eff}}({\bf k})$ is of the same form as the bare hamiltonian, 
the new eigenvalues $\pm \sqrt{\tilde{E}_{{\bf k} \pm}}$ 
can be computed similarly to be,  
\begin{align}
  \tilde{E}_{{\bf k} \pm} = (D_{1}({\bf k}) \pm D_{4}({\bf k}))^2 + 
  (D_{2}({\bf k}) \pm D_{5}({\bf k}))^2 + (D_{3}({\bf k}))^2.  \nonumber
\end{align}
The position of the Weyl nodes can be found by solving for the zeros of 
$\tilde{E}_{{\bf k} \pm}$ i.e., 
\begin{align*}
  D_{1}({\bf k}) = \pm D_{4}({\bf k}) \text{ , } D_{2}({\bf k}) = \pm D_{5}({\bf k}) \text{ , } D_{3}({\bf k}) = 0.
\end{align*}
At $\epsilon = 0$, weak intensity of light ($e^* A \ll 1$) and for $ma < \lambda < (m + 4t_h) a$, there are 4 Weyl nodes at 
$k_x = 0 = k_z$, $\sin(k_y a) = \pm a m_{\text{eff}} / \lambda_{\text{eff}}$ 
where the effective parameters are,
\begin{align}
m_{\text{eff}} &= m + 2t_h \left[ 1 - J_0(e^* A_x) \right] 
\approx m + \frac{t_h}{2}(e^* A_x)^2 + O(A_x^4), \nonumber \\
\lambda_{\text{eff}} &= \lambda J_0(e^* A_y) 
\approx \lambda - \frac{\lambda}{4}(e^* A_y)^2 + O(A_y^4). \nonumber
\end{align}
The positions of Weyl nodes along $k_ya$ axis i.e.,  
$\pm \sin^{-1} ( a m_{\text{eff}} / \lambda_{\text{eff}})$ and 
$\pm \left[ \pi - \sin^{-1} ( a m_{\text{eff}} / \lambda_{\text{eff}}) \right]$ are, (to $O(A^4)$)
\begin{align}\nonumber
\sin^{-1}\left(\frac{a m_{\text{eff}}}{\lambda_{\text{eff}}} \right) \approx 
\sin^{-1}\left(\frac{a m}{\lambda} \right) 
+ \frac{a}{4}\frac{2t_h(e^* A_x)^2 + m (e^* A_y)^2}{\sqrt{\lambda^2 - a^2 m^2}}.
\end{align}
Then the separation between two nearby Weyl nodes is 
$ 2 \tilde{k}_0 a = \pi - 2 \sin^{-1} ( a m_{\text{eff}} / \lambda_{\text{eff}})$ and for 
weak intensities this is, 
\begin{align}\nonumber
2 \tilde{k}_0 a \approx 2 k_0 a - \frac{a}{2}\frac{2t_h(e^* A_x)^2 + m (e^* A_y)^2}{\sqrt{\lambda^2 - a^2 m^2}}.
\end{align}
Using $t_h = 0.5$ and $a = 1 = \hbar$ we get the Eq (5) of the main text. 

\subsection{Green's function method for computing the Josephson current}

\begin{figure}[t]
  \includegraphics[width=0.45\textwidth]{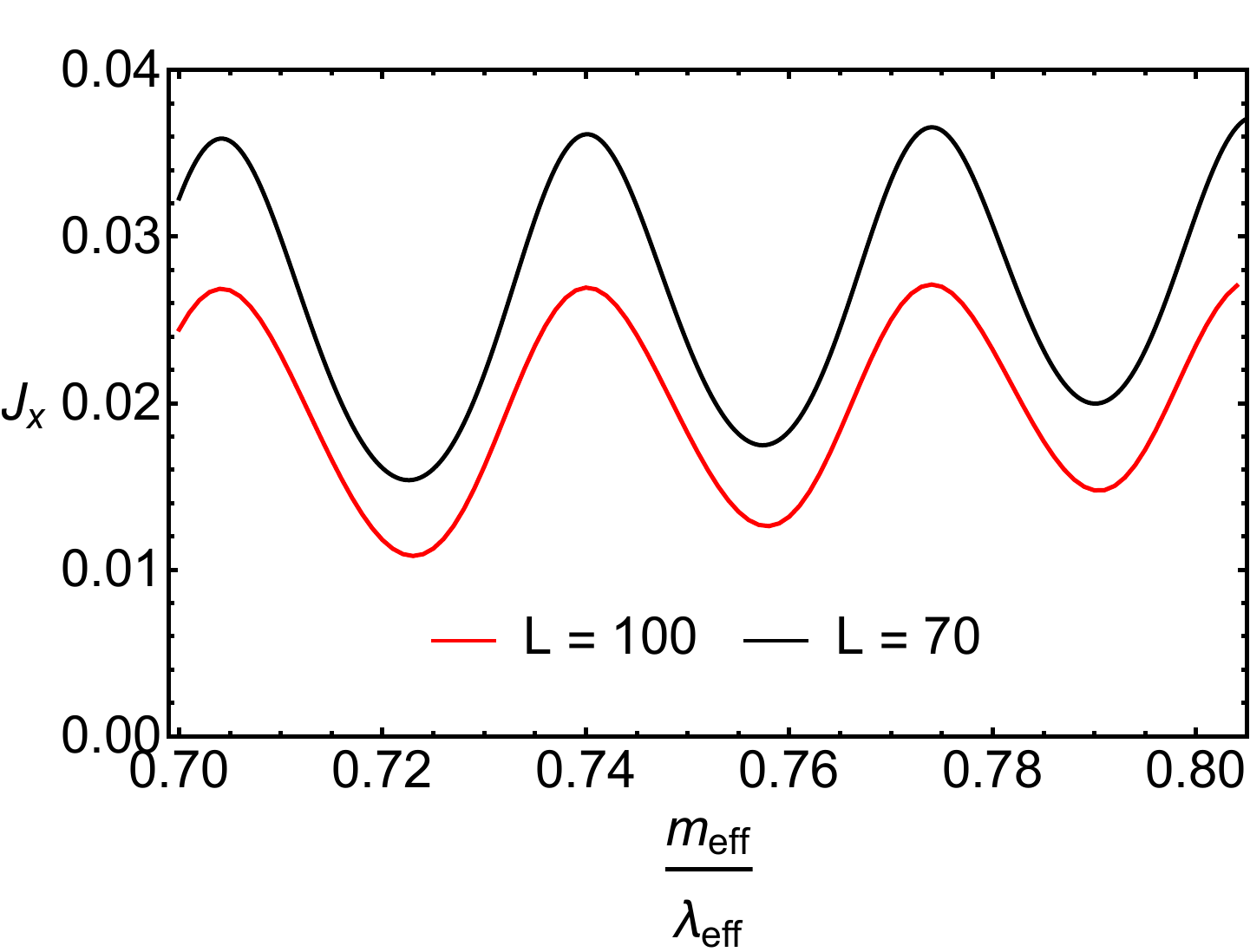}
  \caption{Variation of Josephson current along $x$ direction at phase difference $\phi/2$. The frequency of oscillations is independent of 
  the length $L$ of the WSM showing that these are not $k_0 L$ oscillations. The current $J_x$ is 
  in units of $e \lambda/ \hbar a$. 
  The parameters used are $m = 0.7$, $\lambda = 1.0$, $\hbar = e = a = 1$, 
  $t_h = 0.5$, $\Delta = 0.025$, $\omega = 20$. } \label{fig:Jx}
\end{figure}

To compute the Josephson current along a given direction, we 
write the effective hamiltonian $\hat{\mathcal{H}}_{\text{eff}}$ 
in Nambu basis, as a tight 
binding model with open boundary conditions in that direction 
and with periodic boundary conditions in the two perpendicular directions. 
Then if $\psi_{r,{\bf k}}$ is the eight component electron-hole annihilation operator at 
site $r$ and perpendicular wavenumber ${\bf k}$, we can write - 
$$\hat{\mathcal{H}}_{\text{eff}} = \sum_{\langle r,r'\rangle, {\bf k}} 
\psi^{\dagger}_{r,{\bf k}} H_{\text{hop}} (r-r', {\bf k}) \psi_{r',{\bf k}} 
+ \sum_{r, {\bf k}} \psi^{\dagger}_{r,{\bf k}} H_{\text{d}} (r,{\bf k}) \psi_{r,{\bf k}} $$

To maintain unitarity in the problem, we must have 
$ \left[ H_{\text{hop}} (r-r', {\bf k}) \right]^{\dagger} = H_{\text{hop}} (r'-r, {\bf k}) $
and $ \left[ H_{\text{d}} (r,{\bf k}) \right]^{\dagger} = H_{\text{d}} (r,{\bf k}) $. 
Now $\hat{N}_r = \sum_{\bf k} \psi^{\dagger}_{r,{\bf k}} \psi_{r,{\bf k}}$ 
is the number operator at site $r$ of the system. 
The rate of increase of charge at site $r$ is the sum of currents from 
nearest neighbour sites of $r$ to it, 
\begin{align} \nonumber
  -e \langle \dot{\hat{N}}_r \rangle = J_{r+\delta \rightarrow r} + 
  J_{r-\delta \rightarrow r} = -\frac{ei}{\hbar} [\hat{\mathcal{H}}_{\text{eff}}, \hat{N}_r] . 
\end{align}
In equilibrium (or in a steady state) $\langle -e \dot{\hat{N}}_r \rangle = 0$ but there might be a net
current along one direction ie $J_{r-\delta \rightarrow r} = -J_{r+\delta \rightarrow r}  \neq 0$. 
Only terms in $\hat{\mathcal{H}}_{\text{eff}}$ connecting site $r$ to other sites 
can contribute to the commutator. Thus we find,
\begin{equation} \begin{aligned} \nonumber
J_{r-\delta \rightarrow r} = \frac{ei}{\hbar} &\left[ 
\langle \psi^{\dagger}_{r,{\bf k}} H_{\text{hop}} (\delta, {\bf k}) \psi_{r-\delta,{\bf k}} \rangle - \right. \\
&\left. ~~\langle \psi^{\dagger}_{r-\delta,{\bf k}} H_{\text{hop}} (-\delta, {\bf k}) \psi_{r,{\bf k}} \rangle 
\right]. \end{aligned} \end{equation}
The equal time averages in above equation have to be found through the lesser 
Green's function $G^{+-}(\omega)$. In equilibrium, 
$$ G^{+-}(\omega) = f(\omega) [G^A(\omega) - G^R(\omega)] $$
where $f(\omega)$ is the Fermi-Dirac distribution function and $G^{R(A)}(\omega)$ is the 
retarded (advanced) Green's function of the complete device. The advanced and 
retarded Green's functions are related by 
\begin{align} G^A(\omega) = \left[ G^R(\omega) \right]^{\dagger}~. \end{align}
\noindent
For an isolated (but irradiated) WSM, the retarded Green's function is,
$$
G^R_{0}(\omega) = \left[ (\omega + i \eta) \hat{\mathcal{I}} - \hat{\mathcal{H}}_{\text{eff}} \right]^{-1} .
$$
We model the coupling with the superconductor by 
adding a self-energy $\hat{\Sigma}(\omega)$ \cite{Khanna2014, Khanna2016}
to the first and last sites of the 
bare Green's function $G^R_{0}$. $\hat{\Sigma}$ accounts for 
tunnelling processes between the WSM and superconductor at the boundaries. 
The total retarded Green's function of the SWS device is 
$$
G^R(\omega) = \left[ (\omega + i \eta)  
\hat{\mathcal{I}} -  \hat{\mathcal{H}}_{\text{eff}} - \hat{\Sigma}(\omega) \right]^{-1}.
$$

In this work, we numerically compute the retarded Green's function $G^R(\omega)$ and 
use that to find the lesser Green's function $G^{+-}(\omega)$. Integration over $\omega$ 
yields the equal time averages required to find the Josephson current $J_{r-\delta \rightarrow r}$. 

{\it Along y direction : }
In this case, the effective hamiltonian is written as a tight-binding model with open 
boundary conditions in the $y$ direction.  
Using $\xi$ to denote the particle-hole degree of freedom in Nambu basis and the notations defined earlier this is, 
\begin{align} \nonumber
&H_{\text{hop}} (\hat{y}, {\bf k}) = \frac{i}{2} \frac{\lambda}{a} J_0(e^* A_y) \xi^z \sigma^y 
  - \frac{\epsilon}{2} J_0(e^* A_y) \tau^y \xi^z \sigma^x,
\\ \nonumber
&\begin{aligned} H_{\text{d}} &({\bf k}) = 
  \frac{\lambda}{a} J_0(e^* A_x) \sin (k_x a) \xi^z \sigma^x 
  + \frac{\lambda}{a} \sin (k_z a) \xi^z \sigma^z + \\ 
  &\left[ m + 2t_h \left(2 - J_0(e^* A_x) \cos (k_x a) - \cos (k_z a) \right) \right] \tau^y \xi^z \sigma^y + \\
  &\epsilon[1 - \cos(k_z a)] \tau^y \xi^z \sigma^x.
\end{aligned} \end{align}
Here we have assumed that the incident light is linearly polarized, i.e. $\theta = \pi/2$. 
These expressions can be used to compute the current $J$ which is discussed in the main text. 

{\it Along x direction : }
In this case, the effective hamiltonian is written as a tight-binding model with open 
boundary conditions in the $x$ direction.  
This is given by, 
\begin{align} \nonumber
&H_{\text{hop}} (\hat{x}, {\bf k}) = \frac{i}{2} \frac{\lambda}{a} J_0(e^* A_x) \xi^z \sigma^x 
  - t_h J_0(e^* A_x) \tau^y \xi^z \sigma^y,
\\ \nonumber
&\begin{aligned} H_{\text{d}} &({\bf k}) = 
  \frac{\lambda}{a} J_0(e^* A_y) \sin (k_y a) \xi^z \sigma^y 
  + \frac{\lambda}{a} \sin (k_z a) \xi^z \sigma^z + \\ 
  &\left[ m + 2t_h \left(2 - \cos (k_z a) \right) \right] \tau^y \xi^z \sigma^y + \\
  &\epsilon[1 - J_0(e^* A_y) \cos (k_y a) - \cos(k_z a)] \tau^y \xi^z \sigma^x.
\end{aligned} \end{align}

These expressions can be used to compute the current $J_x$, shown in Fig.~\ref{fig:Jx}. 
The current along $x$ also has oscillations as a function of  $m_{\text{eff}}/\lambda_{\text{eff}}$,  
but no  $0$-$\pi$ transitions. Moreover, 
the frequency of this oscillation is independent of the length $L$ of the WSM. 
Therefore this is not the same $k_0 L$ oscillation that the current along the $y$ direction shows. 
Rather this is due to the changes in the density of states and other details of the model. 
Note that transport along $x$ direction cannot occur at normal incidence because the 
Weyl nodes are at finite $k_y$. 
The $J_x$ shown in Fig.~\ref{fig:Jx} is the total current from all transverse momenta,
whereas the $J$ along $y$ shown in the main text is 
the current at normal incidence $k_x = k_z = 0$.


\begin{thebibliography}{99}

\bibitem{Vishwanath2011} X.~Wan, A.~M.~Turner, A.~Vishwanath and S.~Y.~Savrasov, Phys. Rev. B {\bf 83}, 205101 (2011).

\bibitem{Burkov2011a}
A.~A.~Burkov and L.~Balents,
\newblock \prl\ {\bf 107}, 127205 (2011).


\bibitem{Burkov2011b}
A.~A.~Burkov, M.~D.~Hook and L.~Balents,
\newblock \prb\ {\bf 84}, 235126 (2011).

\bibitem{Zyuzin2012a}
A.~A.~Zyuzin, S.~Wu and A.~A.~Burkov,
\newblock \prb\ {\bf 85}, 165110 (2012).

\bibitem{Hosur2012}
P.~Hosur, S.~A.~Parameswaran and A. Vishwanath, \prl\ {\bf 108}, 046602 (2012).

\bibitem{Nielsen1981}
H.~B.~Nielsen and M.~Ninomiya,
\newblock Phys. Lett. B {\bf 105}, 219 (1981).


\bibitem{Vazifeh2013}
M.~M.~Vazifeh and M.~Franz,
\newblock \prl\ {\bf 111}, 027201 (2013).

\bibitem{Turner2013}
A. M. Turner and A. Vishwanath, arXiv:1301.0330.


\bibitem{Son2013}
D.~T.~Son and B.~Z.~Spivak,
\newblock \prb\ {\bf 88}, 104412 (2013).

\bibitem{Biswas2013}
R.~R.~Biswas and Shinsei Ryu, \prb\ {\bf 89}, 014205 (2014).


\bibitem{Hosur2013}
P.~Hosur and X.~Qi, Comptes Rendus Physique {\bf 14}, 857 (2013).

\bibitem{Burkov2014}
A.~A.~Burkov,
\newblock  \prl\ {\bf 113}, 247203 (2014).


\bibitem{Gorbar2014}
E.~V.~Gorbar, V.~A.~Miransky and I.~A.~Shovkovy,
\newblock \prb\ {\bf 89}, 085126 (2014).

\bibitem{Uchida2014}
S.~Uchida, T.~Habe and Y.~Asano,
\newblock J. Phys. Soc. Jpn. {\bf 83}, 064711 (2014).


\bibitem{Khanna2014}
U.~Khanna, A.~Kundu, S.~Pradhan and S.~Rao, \prb\ {\bf 90}, 195430 (2014).

\bibitem{Ominato2014}
Y.~Ominato and M.~Koshino, \prb\ {\bf 89}, 054202 (2014).

\bibitem{Sbierski2014}
B.~Sbierski, G. Pohl, E.~J.~Bergholtz and P.~W.~Brouwer, 
\newblock \prl\ {\bf 113}, 026602 (2014).

\bibitem{Burkov2015a}
A.~A.~Burkov,
\newblock Journal of Physics: Condensed Matter {\bf 27}, 113201 (2015).

\bibitem{Burkov2015b}
A.~A.~Burkov,
\newblock \prb\ {\bf 91}, 245157 (2015).


\bibitem{Goswami2015}
P.~Goswami, J.~H.~Pixley  and S.~Das Sarma, \prb\ {\bf 92}, 075205 (2015).

\bibitem{Baum2015}
Y.~Baum, E.~Berg, S.~A.~Parameswaran and A.~Stern, Phys. Rev. X {\bf 5}, 041046 (2015).

\bibitem{Khanna2016}
U. Khanna, D. K. Mukherjee, A. Kundu and S. Rao, \prb\ {\bf 93}, 121409(R) (2016).

\bibitem{Behrends2016}
J. Behrends, A. G. Grushin, T. Ojanen and J. H. Bardarson, \prb\ {\bf 93}, 075114 (2016).

\bibitem{Rao2016}
S. ~Rao, arXiv:1603.02821.

\bibitem{Baireuther2016a}
P. Baireuther, J. A. Hutasoit, J. Tworzydlo and C. W. J. Beenakker, New J. Phys. {\bf 18}, 045009 (2016).



\bibitem{Tao2016}
T. Zhou, Y. Gao and Z. D. Wang, \prb\ {\bf 93}, 094517 (2016).


\bibitem{Marra2016}
P. Marra, R. Citro and A. Braggio, \prb\ {\bf 93}, 220507(R) (2016).

\bibitem{Li2016}
X. Li, B. Roy and S. Das Sarma, \prb\ {\bf 94}, 195144 (2016).

\bibitem{Baireuther2016b}
P. Baireuther, J. Tworzydlo, M. Breitkreiz, I. Adagideli and C. W. J. Beenakker, New J. Phys. {\bf 19}, 025006 (2017).


\bibitem{Madsen2016}
K. A. Madsen, E. J. Bergholtz and P. W. Brouwer, Phys. Rev. B {\bf 95}, 064511 (2017).

\bibitem{Xu2015a}
S.-Y.~Xu, I.~Belopolski, N.~Alidoust, M.~Neupane, G.~Bian, C.~Zhang, R.~Sankar, G.~Chang, Z.~Yuan, C.-C.~Lee, S.-M.~Huang, H.~Zheng, J.~Ma, D.~S.~Sanchez, B.~Wang, A.~Bansil, F.~Chou, P.~P.~Shibayev, H.~Lin, S.~Jia, and M.~Z.~Hasan, Science {\bf 349}, 613 (2015).


\bibitem{Xu2015b}
S.-Y.~Xu, N.~Alidoust, I. ~Belopolski, Z. ~Yuan, G.~Bian, T.-R. ~Chang, H. Zheng, V. N. Strocov, D.~S.~Sanchez, G.~Chang, C. Zhang, D. Mou, Y. Wu, L. Huang, C.-C.~Lee, S.-M.~Huang, B.~Wang, A.~Bansil, H.-T. Jeng, T. Neupert, A. Kaminski, H. Lin, S. Jia and M.~Z.~Hasan, Nat. Phys. {\bf 11}, 748 (2015).


\bibitem{Lv2015a}
B.~Q.~Lv, H.~M.~Weng, B.~B.~Fu, X.~P.~Wang, H.~Miao, J.~Ma, P.~Richard, X.~C.~Huang, L.~X.~Zhao, G.~F.~Chen, Z.~Fang, X.~Dai, T.~Qian and H.~Ding,
\newblock Phys. Rev. X {\bf 5}, 031013 (2015).

\bibitem{Lv2015b}
B.~Q.~Lv, N. ~Xu, H.~M.~Weng, J. Z. Ma,  P.~Richard, X.~C.~Huang, L.~X.~Zhao, G.~F.~Chen, C. E. Matt, F. Bisti, V. N. Strocov, J. Mesot, Z.~Fang, X.~Dai, T.~Qian, M. Shi and H.~Ding,
Nat. Phys. {\bf 11}, 724 (2015).


\bibitem{Lu2015}
L. ~Lu, Z. ~Wang, D. ~Ye, L. ~Ran, L. ~Fu, J. ~D. ~Joannopoulos and M.~ Soljacic, Science {\bf 349},  622 (2015).

\bibitem{Jia2016}
S. Jia, S.-Y. Xu and M. Z. Hasan, Nat. Mat. {\bf 15},  1140 (2016).

\bibitem{Wang2014}
R. Wang, B. Wang, R. Shen, L. Sheng and D. Y. Xing, Eur. Phys. Lett. {\bf 105}, 17004 (2014).

\bibitem{Hubener2016}
H. Hubener, M. A. Sentef, U. De Giovannini, A. F. Kemper and A. Rubio, Nat. Comm. {\bf 8}, 13940 (2017).

\bibitem{Ishizuka2016}
H.  Ishizuka, T.  Hayata, M.  Ueda and N.  Nagaosa, Phys. Rev. Lett. {\bf 117}, 216601 (2016).


\bibitem{Chan2016}
C.-K. Chan, Y.-T. Oh, J. H. Han and P. A. Lee, Phys. Rev. B {\bf 94}, 121106(R) (2016).

\bibitem{Yan2016}
Z. Yan and Z. Wang, \prl\ {\bf 117}, 087402 (2016).

\bibitem{Deb2017}
O. Deb and D. Sen, arXiv: 1701.03661.


\bibitem{Chen2016}
A. Chen and M. Franz, \prb\ {\bf 93}, 201105(R) (2016).


\bibitem{Bulaevski1977} L. N. Bulaevskii, V. V. Kuzii and A. A. Sobyanin, JETP Lett.  {\bf 25}, 290 (1977).

\bibitem{Buzdin1982} A. I. Buzdin,  L. N. Bulaevskii and S. V. Panyukov, JETP Lett. {\bf 35}, 178 (1982).

\bibitem{Buzdin1991} A. I. Buzdin and M. Y. Kupriyanov, JETP Lett. {\bf 53}, 321 (1991).

\bibitem{Rmpsfs} For reviews, see A. A. Golubov, M. Y. Kupriyanov and  E. Il'ichev, Rev. Mod. Phys. {\bf 76}, 411 (2004);
A. I. Buzdin,  Rev. Mod. Phys. {\bf 77}, 935 (2005); F. S. Bergeret, A. F. Volkov and K. B. Efetov,  Rev. Mod. Phys. {\bf 77}, 1321 (2005).

\bibitem{Ryazanov2001} V.~V.~Ryazanov, V.~A.~Oboznov, A.~Yu.~Rusanov, A.~V.~Veretennikov, A.~A.~Golubov and J.~Aarts, 
\prl\ {\bf 86}, 2427 (2001).

\bibitem{supp}
See supplementary materials for further details.


\bibitem{Mikami2016}
T.~Mikami, S.~Kitamura, K.~Yasuda, N.~Tsuji, T.~Oka and H. Aoki,
\newblock \prb\ {\bf 93}, 144307 (2016).


\bibitem{Feldman1984}
E.~B.~Fel'dman,
\newblock Phys. Lett. A {\bf 104}, 479 (1984).

\bibitem{Mananga2011}
E.~S.~Mananga and T.~Charpentier,
\newblock J. Chem. Phys. {\bf 135}, 044109 (2011).

\bibitem{Casas2001}
F.~Casas, J.~A.~Oteo and J.~Ros,
\newblock J. Phys. A: Math. Gen. {\bf 34}, 3379 (2001).


\bibitem{Kuwahara2016}
T.~Kuwahara, T.~Mori and K.~Saito,
\newblock Annals of Physics {\bf 367}, 96-124 (2016).

\bibitem{Eckardt2015}
A.~Eckardt and E.~Anisimovas,
\newblock New J. Phys. {\bf 17}, 093039 (2015).


\bibitem{Bukov2015}
M.~Bukov, L.~D'Alessio, and A.~Polkovnikov,
\newblock Advances in Physics {\bf 64}, No. 2, 139-226 (2015).


\bibitem{Vidal1994}
A.~Martin-Rodero, F.J.~Garcia-Vidal and A.~Levy Yeyati,
\newblock Phys. Rev. Lett. {\bf 72}, 554 (1994).

\bibitem{Beenakker1704}
N. Bovenzi, M. Breitkreiz, P. Baireuther, T. E. O'Brien, J. Tworzydlo, I. Adagideli and C. W. J. Beenakker, arXiv: 1704.02838.




\end{thebibliography}
\end{document}